\documentclass[journal]{IEEEtran}
\makeatletter

\newcommand{\Rmnum}[1]{\expandafter\@slowromancap\romannumeral #1@}
\makeatother
\usepackage{times}
\usepackage{balance}
\usepackage{graphicx}
\usepackage{enumerate}
\usepackage{amsfonts}
\usepackage{subfigure}
\usepackage{cite}
\usepackage{multirow}
\usepackage{amstext}
\usepackage{bm}
\usepackage{amsmath,graphicx}
\usepackage{amsmath,bm}
\unboldmath

\newenvironment{packed_enum}{
\begin{enumerate}[\indent 1)]
  \setlength{\itemsep}{1pt}
  \setlength{\parskip}{0pt}
  \setlength{\parsep}{0pt}
}{\end{enumerate}}

\unboldmath

\begin{document}

\title{Data hiding in Fingerprint Minutiae Template for Privacy Protection}

\author{Sheng Li, Xin Chen, Zhigao Zheng, Zhenxin Qian, Guorui Feng, Xinpeng Zhang}
\maketitle

\boldmath
\begin{abstract}
In this paper, we propose a novel scheme for data hiding in the fingerprint minutiae template, which is the most popular in fingerprint recognition systems. Various strategies are proposed in data embedding in order to maintain the accuracy of fingerprint recognition as well as the undetectability of data hiding. In bits replacement based data embedding, we replace the last few bits of each element of the original minutiae template with the data to be hidden. This strategy can be further improved using an optimized bits replacement based data embedding, which is able to minimize the impact of data hiding on the performance of fingerprint recognition. The third strategy is an order preserving mechanism which is proposed to reduce the detectability of data hiding. By using such a mechanism, it would be difficult for the attacker to differentiate the minutiae template with hidden data from the original minutiae templates. The experimental results show that the proposed data hiding scheme achieves sufficient capacity for hiding common personal data, where the accuracy of fingerprint recognition is acceptable after the data hiding.
\end{abstract}


\IEEEpeerreviewmaketitle

\section{Introduction}
Nowadays, biometrics such as fingerprint, face and iris are widely used in authentication systems. Usually, biometrics need to be stored in a database for subsequent authentication. However, templates stored in the database are at the risk of being stolen or modified. Once the templates are stolen, it is difficult to be replaced like passwords. Meanwhile, the user's private information (e.g., name, address, bank account number and etc.) associated with the stolen template would also be leaked. Thus, it is of paramount importance to protect the privacy of biometrics data.

The simplest way to protect the biometric template is using cryptography that aims at encrypting data into some information meaningless to unauthorized parties. The encrypted template is protected as long as the decryption key is secure. However, cryptography is not the best way to protect the biometric template due to the non-smooth property of the encryption functions, which cannot handle the intra-class variability. Any little difference in the biometric template could make a significant change in the encrypted template. Thus, it is not possible to perform the biometric matching directly in the encrypted domain, a correct decryption of the encrypted template is necessary before every matching attempt.

In recent years, significant efforts have been put in developing specific protection techniques for biometric template, which can be classified into three major categories including:
\begin{packed_enum}
\item \textbf{Cancelable biometrics generation}: transforms a biometric signal or template to another domain using a transformation function. The biometric matching can be performed directly on the transformed templates with little or no degradation of the matching performance. The transformation can be further classified into non-invertible transformation \cite{ratha07,lee07,Ferrara2012} and biohashing \cite{jin04} based on the property of the transformation function. In non-invertible transformation, a key or password is used to guide the transformation function. In biohashing, the transformation function depends on an user specific pseudo-random projection matrix.
\item \textbf{Biometric cryptosystem}: aims at creating a cryptographic key for the biometric template. In biometric cryptosystem, some helper data is stored in association with the user identity. The helper data is usually a combination of the template and a key as well as some other information, which should not expose any important information of the template. It should be difficult to recover the original biometric template or extract the key from the helper data. Biometric matching is carried out by checking the validity of the key extracted from the query biometric data with the corresponding helper data. The biometric cryptosystem can be further classified into fuzzy commitment \cite{Teoh07} and fuzzy vault \cite{nandakumar07,Yang2014Hutifs,li16} based on whether the technique is able to handle point set features.
\item \textbf{Biometrics Data hiding}: conceals personal data on a cover biometric template \cite{jain2003hiding,li10a}. To hide the secret into the template, one needs to imperceptibly alter the value of the pixels in the template. While the location of these pixels is usually dependent on a key. In such a way, the secret and the biometric template is combined together, which still appears as an image. The template with hidden secret is visually similar as the original template, so it can be used directly for the biometric matching.
\end{packed_enum}
Despite the relative high security provided by the cancelable biometrics generation and the biometric cryptosystem, they inevitably destroy the structure of the biometrics template. The protected template generated by these two types of schemes are noise like, which can be easily differentiated from the original templates. This may lead to the attacker's interest for cracking the protected template \cite{ross11a,Ratha04}. For biometrics data hiding, however, it is able to preserve the main feature of the biometrics template. The protected template with hidden secret is difficult to be differentiated from the original templates.

People have proposed various biometrics data hiding techniques in literature, most of which are designed for image level biometrics templates including the color image, grayscale image and the binary image. In this paper, we focus on developing data hiding techniques to protect the privacy of fingerprint templates. Instead of working on the fingerprint images, we investigate the possibilities to hide data in the fingerprint minutiae (termed as minutiae for short) which is the most popular fingerprint feature. Concretely, three strategies are proposed for data hiding in minutiae template. In the first strategy, we simply alter the minutiae locations and directions by replacing their last few bits with the data to be embedded. In the second strategy, the alteration is performed such that its impact on the fingerprint recognition accuracy is minimized. To reduce the detectablity of data hiding, we also propose an order preserving mechanism to maintain the order of the minutiae points stored in the minutiae template, so that it is difficult for the attacker to differentiate our protected minutiae template from the original minutiae templates.

The rest of the paper is organized as follows. In Section \ref{background}, we introduce the related work regarding the fingerprint data hiding. In Section \ref{minutiae}, we briefly explain the format of the minutiae template. The details of our proposed method for data hiding in the minutiae templates are elaborated in Section \ref{proposedMethod}, followed by the experimental results and conclusions in Section \ref{experiment} and Section \ref{conclusions}.

\section{Related work}
\label{background}
In literature, various data hiding approaches are proposed for the protection of fingerprint templates.  Ratha \emph{et al}. \cite{Ratha00} propose a method to hide data in Wavelet Scale Quantization (WSQ) compressed fingerprint image. In data embedding, some high frequency quantized coefficients are chosen randomly based on a seed related to the image content. The least significant bit (LSB) of a chosen coefficient is forced to be equal to the hidden bit. In data extraction, the same seed is determined from the content of the image. Data is extracted according to the LSBs of the randomly chosen high frequency quantized coefficients. An improvement of this method is given in \cite{Ratha04}, where the random seed not only relates to the image content but also the knowledge (i.e., password). So even the attacker knows the data hiding method, he is not able to properly hide the message. In \cite{Ratha04}, the authors also point out that a secure system should not give the impression to the attacker that the system is using a data hiding technique. Thus, they randomly change the LSBs of the coefficients which are not used in data embedding, so as to reduce the detectability of the data hiding technique used in the system. Such an operation will make the statistics of the stego-fingerprint image (i.e., the fingerprint image with hidden data) nearly the same as an original fingerprint image, so the possibility for the attacker to detect the existence of secret data is reduced.

Gunsel \emph{et al}. \cite{Gunsel02} introduce two data hiding schemes for grayscale fingerprint image in spatial domain. The first method inserts a watermark into the fingerprint image after feature extraction, where watermark embedding does not modify the pixels belong to the fingerprint feature region such as bifurcation or endpoint areas. The second method embeds data into the fingerprint image before feature extraction. In this method, watermark embedding is performed by preserving the gradient orientation near a watermark embedding location, i.e., the change of gradient orientation is within a certain range during the watermark embedding. These two schemes are not practically useful due to the extremely low data hiding capacity. For a 256$\times$256 grayscale fingerprint image, the capacity of the first method and the second method is 156 bits and 12 bits, respectively.

Based on what has been done in \cite{Gunsel02}, Jain \emph{et al}. \cite{Jain02, Jain03} propose to apply data hiding techniques on biometrics for two scenarios, i.e., secure communication and biometrics authenticity enhancement. In scenario 1, the fingerprint minutiae that requires to be transmitted is hidden into a cover image for secure communication, where the cover image could be any image. In scenario 2, the eigen-face coefficients are hidden into a grayscale fingerprint image for enhancing the authenticity of the fingerprint and face. During the enrollment, the eigen-face coefficients are extracted from the user's face image and hidden into his grayscale fingerprint image. The grayscale fingerprint image with the hidden data will be stored in the system database or on a smart card. In the authentication, both the face and fingerprint images of the user are needed for query. The system extracts the hidden eigen-face coefficients from the stored fingerprint image. The authentication is successful if the extracted eigen-face coefficients can be matched with the query face image, meanwhile, the stored fingerprint image has to be matched with the query fingerprint. To preserve the crucial features of the fingerprint, the value of the pixels in fingerprint minutiae area or ridge region are fixed during data embedding. However, the fingerprint matching accuracy is still slightly degraded after data embedding.

Motivated by \cite{Jain02, Jain03}, a lot of data hiding schemes are proposed to embed some additional biometric data into the fingerprint image to increase the security or protect the integrity of the latter \cite{Vatsa06,Noore07,Ahmed08}. Vatsa \emph{et al}. \cite{Vatsa06} propose to embed facial features into a fingerprint image in Discrete Wavelet Transform (DWT) domain to protect the face and fingerprint from tempering. Data is embedded in the second LSBs of some selected DWT coefficients. This method provides some robustness against geometric and frequency attacks. However, the fingerprint matching accuracy before and after data embedding are not reported. Noore \emph{et al}. \cite{Noore07} propose to embed both the demographic text data and the face features into a fingerprint for protecting the integrity of the fingerprint image. The demographic text data contains common personal information such as name, ID and gender. In the proposed data hiding scheme, data is embedded in the selected fingerprint texture regions in DWT domain. As the texture regions may not be the same before and after data embedding, 100\% extraction accuracy is not guaranteed in this scheme. In \cite{Ahmed08}, a phase-encoding-based watermarking scheme for protecting the fingerprint is proposed. Some one-dimensional Fourier phase features are extracted and hidden in the fingerprint itself to achieve self-authentication. Data embedding is conducted on the Fourier phase domain by a phase-shift key modulation based spread spectrum embedder. The capacity of this scheme is quite limited, experimental results show that only 30 bits can be hidden into a 300$\times$480 grayscale fingerprint image to guarantee 100\% watermark extraction accuracy.

The above mentioned techniques are designed for the compressed or raw grayscale fingerprint images, which do not work for binary fingerprint images. In \cite{li10} and \cite{li10a}, the authors develop two data hiding schemes for the thinned fingerprint template, which is a binary fingerprint image after thinning. The thinned fingerprint maintains all the key features of a fingerprint and is much smaller in file size compared with its grayscale counterpart. In addition, extracting minutiae features from a thinned fingerprint is much faster. The method proposed \cite{li10} is a lossless data hiding scheme which is capable of reconstructing the original thinned fingerprint after data embedding. In this scheme, data is hidden inside the thinned fingerprint by adding some boundary pixels in the thinned fingerprint. The newly produced boundary pixels can be identified and removed to recover the original thinned fingerprint, so that the fingerprint matching accuracy is not affected after data embedding. In \cite{li10a}, a steganographic scheme is proposed for the thinned fingerprint template, which is able to reduce the detectablity of data hiding technique. In this scheme, data is embedded by exchanging suitable pixels in the thinned fingerprint template according to some rules. No boundary pixels will be produced in such embedding process. Thus, it is difficult for the attacker to identify the thinned fingerprint template with hidden data.

All the existing fingerprint data hiding schemes are developed for fingerprint templates in the image level (i.e., grayscale or binary images). As a matter of fact, most of the fingerprint recognition systems only store fingerprint features for the sake of storage and speed. It is necessary and important to develop suitable data hiding techniques for the feature level fingerprint template. To the best of our knowledge, we are unable to disclose any literature that is developed for data hiding in fingerprint templates in the feature level. In this paper, we propose a data hiding scheme for the fingerprint minutiae template which is the most popular fingerprint features. Different mechanisms are incorporated such that the impact of the data embedding on the fingerprint recognition is minimized and the the detectablity of data hiding is reduced.

\section{The fingerprint minutiae}
\label{minutiae}
\begin{figure}[t]
  \centering
  \includegraphics[scale=0.6]{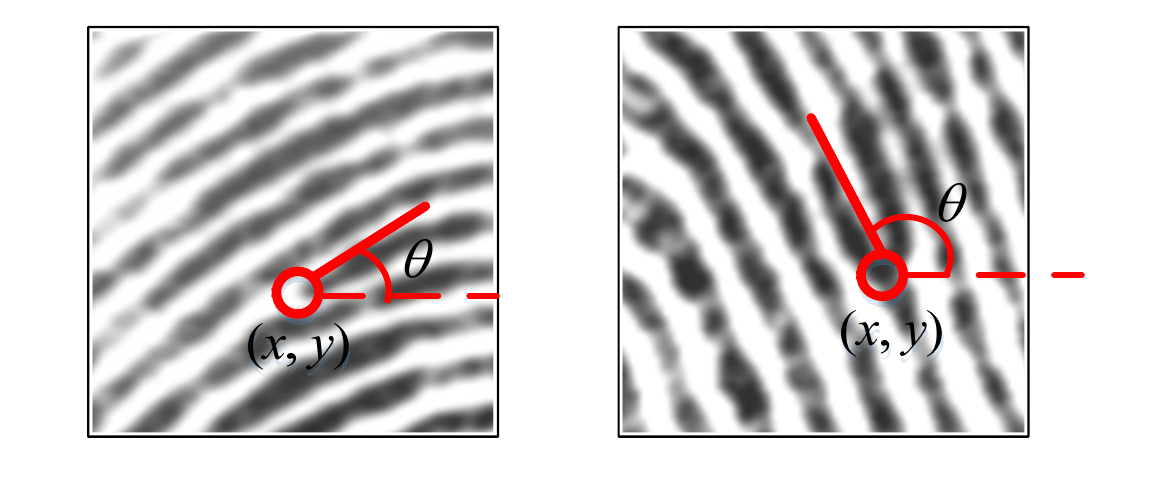}
\caption{A ridge ending (left) and a bifurcation (right) extracted from a real fingerprint. $(x,y)$ refers to the horizontal and vertical coordinate of the minutiae, and $\theta$ is the direction of the minutiae.}
\label{fig_minutiae}
\end{figure}

\begin{figure}[t]
\centering
  \includegraphics[scale=0.5]{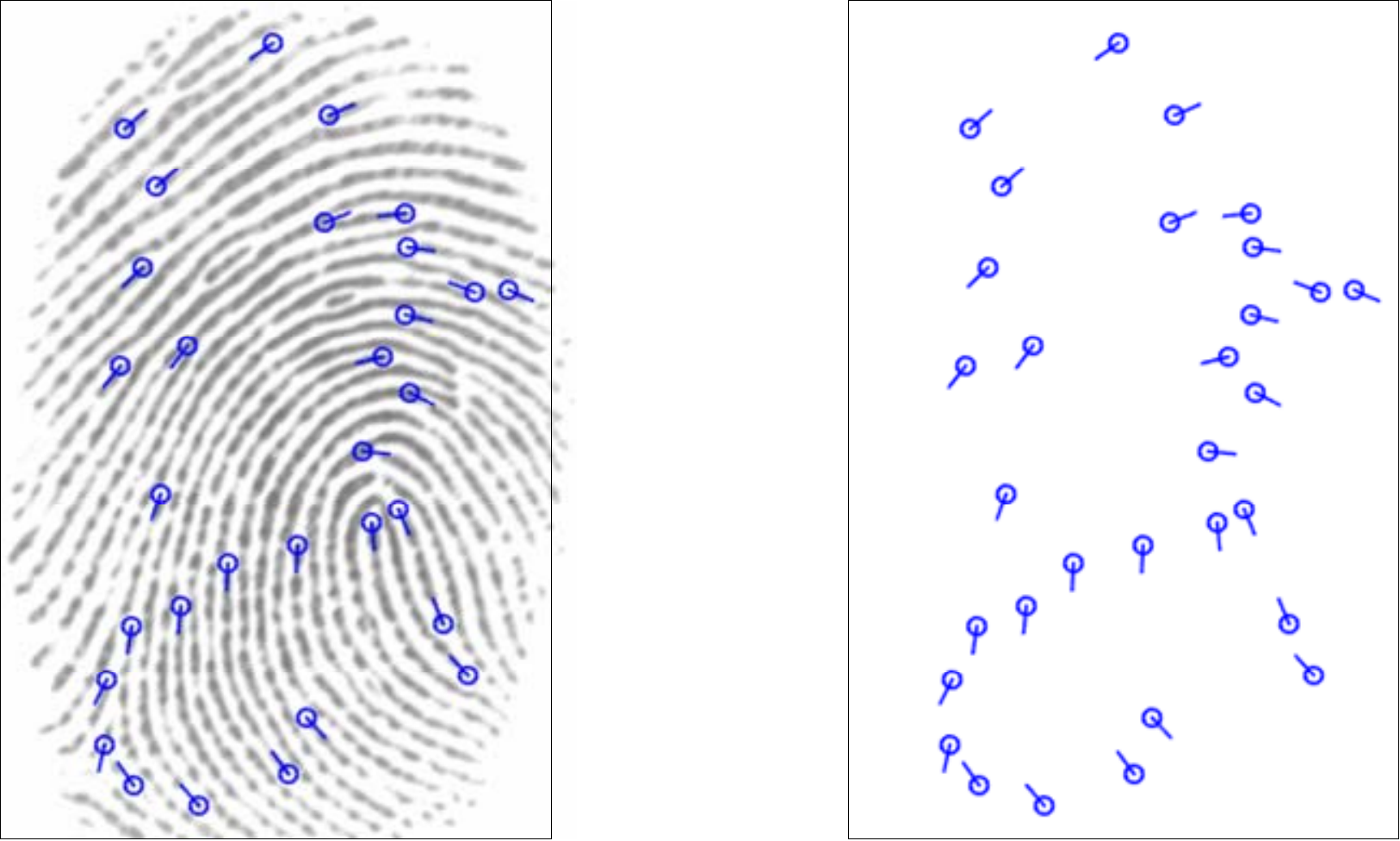}
  \caption{Illustration of a minutiae template extract from a full fingerprint image. Left: the fingerprint image with the minutiae overlapped, right: the minutiae template.}
  \label{fig_template}
\end{figure}

The fingerprint minutiae records the detailed features of a fingerprint. It represents various ways that the fingerprint ridges meets or ends. Typically, a ridge can suddenly come to an end to get a ridge ending, or be divided into two ridges which results to a bifurcation. The fingerprint minutiae are the most popular features in fingerprint recognition. The American National Standards Institute \cite{nist} classifies the minutiae into four types including the ridge ending, the ridge bifurcations, compound (trifurcation or crossovers), and type undetermined. The FBI minutiae-coordinate model \cite{McCabe03} considers only ridge endings and bifurcations, which is widely adopted in the minutiae template. In this model, each minutiae is represented in three fields including
\begin{packed_enum}
\item The x-coordinate (horizontal) of the location.
\item The y-coordinate (vertical) of the location.
\item The minutiae direction which is angle between the tangent to the ridge line at the minutiae and the horizontal axis.
\end{packed_enum}
Fig. \ref{fig_minutiae} shows a ridge ending and a bifurcation of a real fingerprint. An illustration of the minutiae template extracted from a full fingerprint image is shown in Fig. \ref{fig_template}.

In this paper, the minutiae template are generated based on the FBI minutiae-coordinate model. Let's denote the minutiae of a fingerprint images as $\mathbf{M}=\{(x_i, y_i, \theta{_{i}}), 1\leq i\leq N\}$, where $N$ is the number of minutiae points, $(x_i, y_i)$ is the coordinate (i.e., location) of each minutiae point and $\theta_{i}$ is the corresponding direction. In general, a minutia template stores all these minutiae based on the ascending order of $x_i$, as shown in Table \Rmnum{1}.

\begin{table}[t]
\label{table_minutiae}
\setcounter{table}{0}
\caption{The data stored in a minutiae template}
\begin{tabular}{c|c|c|c}
  \hline
   index & x-coordinate & y-coordinate & direction (in degrees)\\
   \hline
  1 & 43 & 152 & 236 \\
  \hline
  2 & 43 & 185 & 236 \\
  \hline
  3 & 46 & 141 & 225 \\
  \hline
  4 & 46 & 125 & 214 \\
  \hline
  5 & 47 & 114 & 56 \\
 \hline
  6 & 48 & 229 & 225 \\
  \hline
    ... &... &... &...\\
  \hline
\end{tabular}
\end{table}

\section{The Proposed Method}
\label{proposedMethod}
\subsection{The system}
\begin{figure*}[t]
\centering
  \includegraphics[scale=0.5]{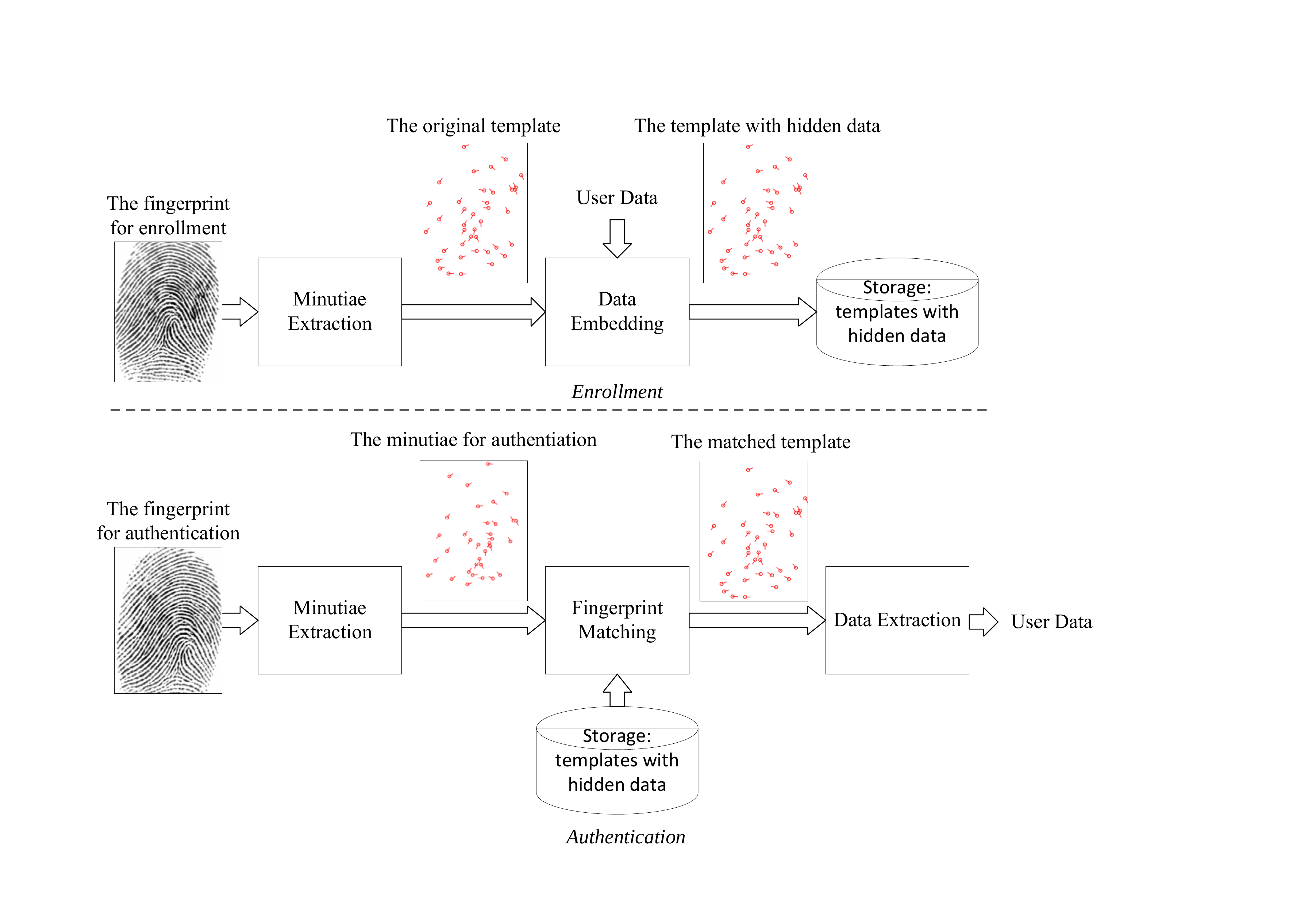}
  \caption{The minutiae based fingerprint recognition system incorporating data hiding.}
  \label{fig_system}
\end{figure*}

Motivate by the system proposed in \cite{Jain03,li10a}, we introduce here a minutiae based fingerprint recognition system which incorporates data hiding for the privacy protection. Fig. \ref{fig_system} illustrates the block diagram of the system. During the enrollment, the user's identity or other private information are embedded into his minutiae template using our proposed scheme. The template with hidden information are then stored in a database. In authentication, an input fingerprint will be matched against all the minutiae templates stored in the database. If the fingerprint matching is successful, the hidden private user information will be extracted from the matched minutiae template for further process or other applications. In such a system, when the database is stolen, the attacker will only have a bunch of minutiae template without any other useful information associated. It would be difficult for the attacker to illegally use the stolen minutiae templates.
\subsection{Data embedding}
\subsubsection{Bits replacement}
We consider a cover minutiae template which contains three different fields including the x-coordinate, y-coordinate, and the direction. Each field has $N$ cover elements with $N$ as the number of minutiae points. The bits replacement based data embedding is performed field by field, where each field is treated equally. For simplicity, we denote $g_i$ as the $i$th of the $N$ cover elements in a certain field. During the data embedding, we embed $b$ secret bits into each cover element $g_i$ by replacing the least $b$ significant bits of $g_i$ with the secret bits. The element with hidden data can be computed as
\begin{equation}
\label{bit_replacement}
z_i=2^{b}\left \lfloor \frac{g_{i}}{2^{b}} \right \rfloor+d\quad ,
\end{equation}
where $d=d_0\times 2^{0}+d_{1}\times 2^{1}+ ... + d_{b}\times 2^{b}$ are the decimal value of the $b$-bits secret information.

\subsubsection{Optimized bits replacement}
The problem of the bits replacement based data embedding is that it changes the locations and directions too much when multi-bits (i.e., $b>1$) secret is embedded into each element. This will significantly reduce the fingerprint recognition accuracy after the data embedding. To mitigate this issue, we propose a optimized bits replacement data embedding scheme in this section.

Given $b$ bits of secret (in terms of a decimal vale of $d$) to be embedded into the cover element $g_i$, we first compute a pair of embedding parameters $p$ and $q$, where
\begin{equation}\label{p_value}
  p=|g_{i}-[2^{b}\left \lfloor \frac{g_{i}}{2^{b}} \right \rfloor-(2^{b}-d)]|
\end{equation}
and
\begin{equation}\label{q_vale}
 q=|g_{i}-(2^{b}\left \lfloor \frac{g_{i}}{2^{b}} \right \rfloor+d)|
\end{equation}
The element after data embedding is then computed as
\begin{equation}
\label{optimized_replacement}
z_{i}= \left\{ \begin{array}{l}
 2^{b}\left \lfloor \frac{g_{i}}{2^{b}} \right \rfloor+d,\quad \quad \quad \quad \; if\ p\geq q \\
 2^{b}\left \lfloor \frac{g_{i}}{2^{b}} \right \rfloor-(2^{b}-d),\quad if\  p< q \\
 \end{array} .\right.
\end{equation}
This algorithm takes into account two strategies of the bits replacement for the cover element. For example, when embedding the secret bits ``11" (i.e., $d$=3) into the cover element ``1100" (i.e., $g_i$ in the form of bits), there are following two strategies to modify $g_i$:
\begin{packed_enum}
\item Change $g_i$ to ``1111" by adding $d=3$;
\item Change $g_i$ to ``1011" by subtracting $2^2-d=1$.
\end{packed_enum}
For the first strategy (i.e., the bits replacement), the difference between the cover element before and after the data embedding is 3, which is only 1 for the second strategy. In our optimized bits replacement, we determine the suitable data embedding strategy for each single element based on the data to be embedded. This is able to reduce the impact of data embedding on fingerprint recognition.

\subsubsection{Order preserving}
As what we have mentioned in Section \ref{minutiae}, the minutiae template usually stores each minutiae points according to the ascending order of the x-coordinate. This may not be the case after the data embedding, where the order of x-coordinates may be disturbed due to the modification of the x-coordinates. Let take the 4th and 5th minutiae points stored in the minutiae template given in Table \ref{table_minutiae} as an example. We embed secret ``10" into the x-coordinate of the 4th minutiae point and secret ``01" into the x-coordinate of the 5th minutiae point. According to Eq. (\ref{optimized_replacement}), the x-coordinate of the 4th and 5th minutiae points will be changed to 46 and 45 after such data embedding. This creates the anomaly in the minutiae templates with hidden data. Such anomaly can be easily noticed and detected by the attacker.

To reduce the detectablity of the data hiding, we propose in this section an order preserving mechanism during the data embedding. For each element, if the inequality
\begin{equation}
 z_i \geq z_{i-1},
\end{equation}
holds after the data embedding using Eq. (\ref{optimized_replacement}), we proceeds to the next element for data embedding. Otherwise, we revise $z_i$ into
\begin{equation}
\label{order_preserving}
  z'_{i}= z_{i}+l2^b,
\end{equation}
where $l\geq1$ is an integer which is computed by
\begin{equation}
\label{integer_l}
 l=\mathop {\arg \min }\limits_{l}(z'_i-z_i).
\end{equation}
This mechanism preserves the order of the x-coordinates of the minutiae template before and after the data embedding. It would be difficult for an attacker to differentiate the protected minutiae template (i.e., the one with hidden data) from the original minutiae templates.

The data embedding steps can be summarized as follows by taking account all the possible strategies.
\begin{packed_enum}
\item Choose one element from each field of the minutiae template sequentially.
\item Embed $b$ secret bits into the element using the bits replacement based data embedding (see Eq. (\ref{bit_replacement})) or the optimized bits replacement based data embedding (Eq. (\ref{optimized_replacement})).
\item Adopt the order preserving mechanism when necessary.
\item Proceed to the next element until all the elements are processed.
\end{packed_enum}

\subsection{Date extraction}
Regardless of the strategies incorporated in data embedding, the secret data can be extracted based on the last $b$ bits of each element from the minutiae template with hidden data, the steps of which can be formulated below.
\begin{packed_enum}
\item Choose one element from each field of the minutiae template based on the same sequence as that of the data embedding.
\item Extract the last $b$ bits of the value of the element as the secret bits.
\item Proceed to the next element until all the elements are processed.
\item Concatenate all the secret bits extracted from each element.
\end{packed_enum}

\section{Experimental results and discussions}
\label{experiment}
Our experiments are evaluated on the FVC2002 DB1\_a database, which contains 800 grayscale fingerprint images from 100 fingers with 8 impressions per finger. The algorithms (MINDTCT and BOZORTH3) provided by the NIST Biometric Image Software (NBIS) \cite{nist10} are used for minutiae extraction and matching. We extract 800 original minutiae templates from the database using MINDTCT with the boundary minutiae points removed. We perform the proposed data embedding into these minutiae templates with different settings of $b$. For simplicity, the templates with hidden data are termed as the protected template. For each protected template, the amount of data that is hidden is $3bN$ bits, where $N$ refers to the number of minutiae points. Thus, we obtain different sets of protected templates corresponding to different $b$, each of which contains 800 protected templates.

To evaluate the impact of data hiding on the fingerprint recognition accuracy. We treat the protected template as the template stored in the database. The original minutiae templates extracted from the other impressions of the same finger are treated as the minutiae for authentication. In total, there are seven such original template for each finger. To compute the false rejection rate (FRR), we match each of the protected templates against the other seven original impressions (templates) of the same finger. This gives us $(8\times7)\times 100 = 5600$ genuine matches. To compute the false rejection rate (FRR), the 100 first impressions of the protected templates are matched against each other, which produces $(100\times99)/2 = 4950$ imposter matches.

\begin{figure}[t]
  \centering
  \fbox{\includegraphics[scale=0.35]{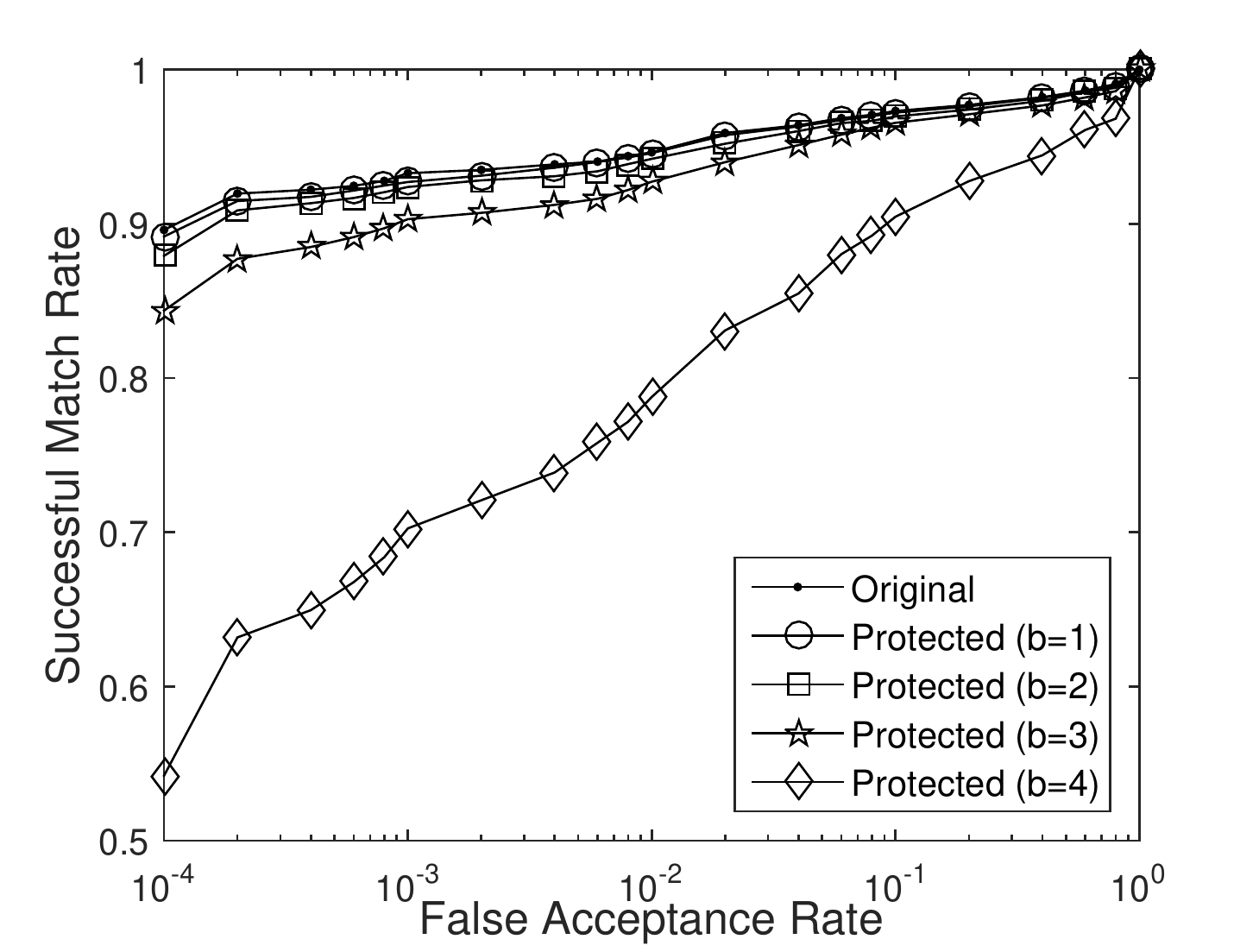}}\\
  \centering
  \caption{The performance of fingerprint recognition with and without data hiding.}
  \label{Fig_performance}
\end{figure}

The performance of the fingerprint recognition for different sets of protected fingerprints as well as the original minutiae template are shown in Fig. \ref{Fig_performance}, where the successful match rate equals to 1$-$FRR and the protected templates are generated based on the optimized bits replacement and order preserving. It can be seen that when $b=4$ bits are embedded, which indicates that 4 bits are embedded into each element of the minutiae template, the successful match rate drops seriously. In general, the accuracy of the fingerprint recognition remains high when $b<4$.

\begin{figure}[t]
  \centering
  \fbox{\includegraphics[scale=0.35]{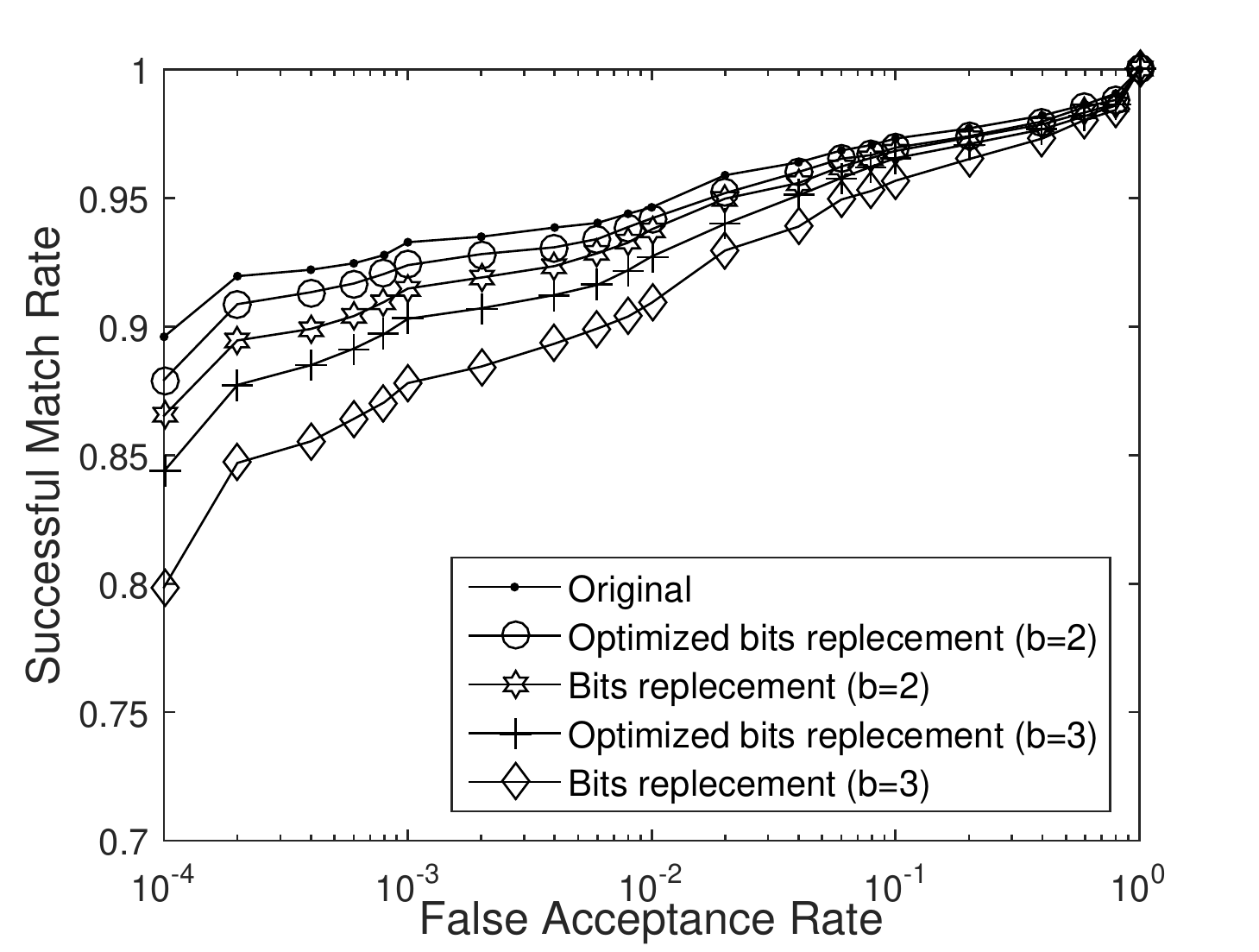}}\\
  \caption{The performance of fingerprint recognition after bits replacement based data hiding and optimized bits replacement based data hiding.}
  \label{Fig_compare}
\end{figure}

\begin{figure*}[t]
  \centering
  \includegraphics[scale=0.5]{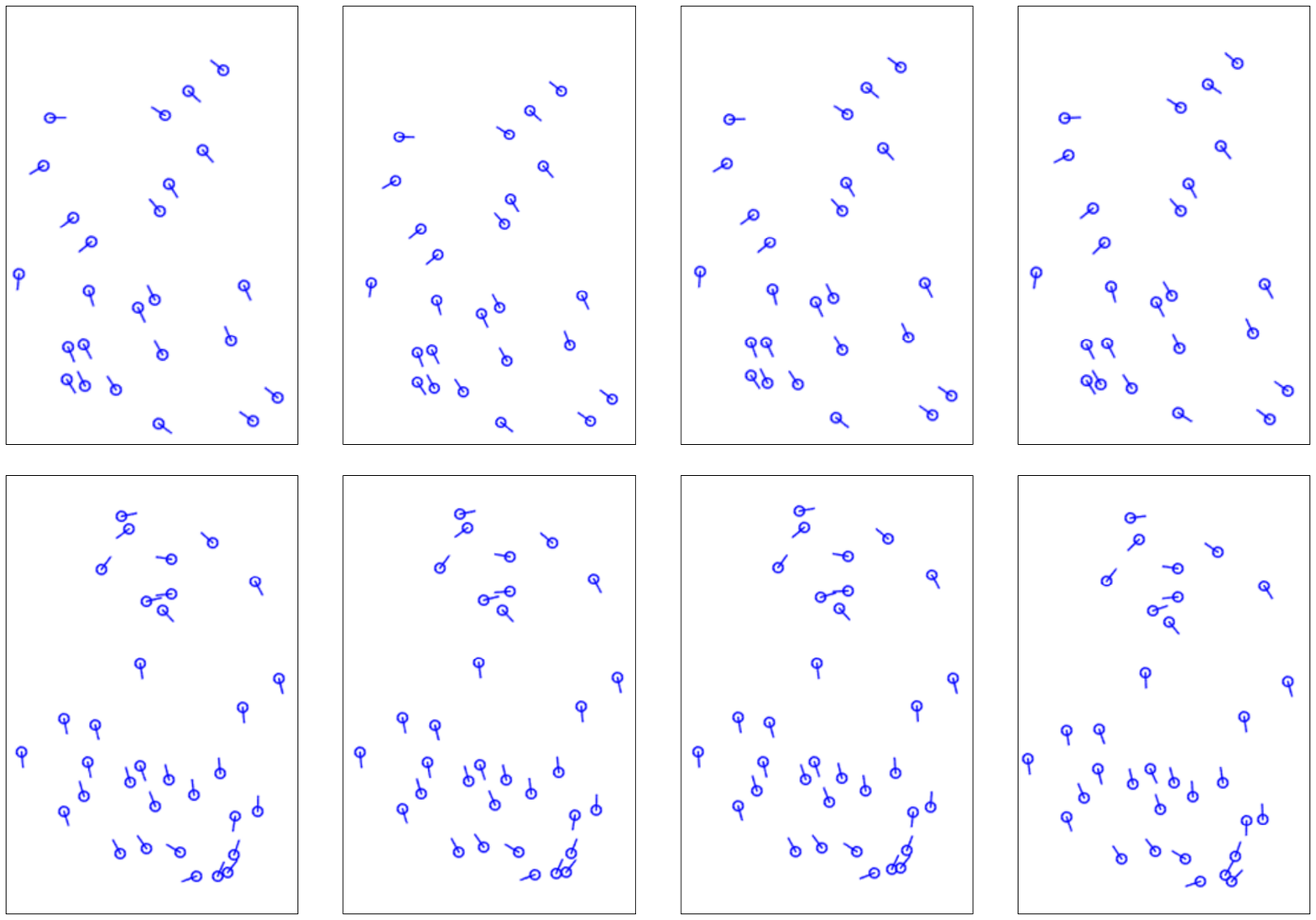}\\
  \caption{The minutiae templates before and after the data hiding. From left to right: the original minutiae template; the protected template ($b=1$); the protected template ($b=2$); the protected template ($b=3$). All the protected templates are generated based data hiding scheme incorporating the optimized bits replacement and order preserving mechanism. }
  \label{Fig_examples}
\end{figure*}

To verity the effectiveness of the optimized bits replacement, we further compare the accuracy of the fingerprint recognition using the optimized bits replacement and the bits replacement in Fig. \ref{Fig_compare}, where the order preserving mechanism is applied for both schemes. Obviously, the optimized bits replacement performs better in terms of the fingerprint recognition. Generally speaking, when more bits are embedded into one element, more improvement will be gained using the optimized bits replacement based data hiding with respect to that using bits replacement based data hiding. At FAR$=10^{-4}$ and $b=2$, the successful match rate using the optimized bits replacement based data hiding is about 1\% higher than that using the bits replacement. This will be increased to over 5\% when $b=3$. Fig. \ref{Fig_examples} lists some examples of the original minutiae templates and the corresponding protected templates obtained based on the optimized bits replacement and order preserving. It can be seen that the protected templates are visually similar to the original minutiae template, which are difficult to be differentiated.

The capacity of our proposed data hiding scheme varies among different minutiae templates. A minutiae template with more minutiae points available is able to hide more secret bits. The statistics of the number of minutiae points of each finger in FVC2002 DB1\_a is given in Table \Rmnum{2}. If we set $b=3$ for the data embedding, the average capacity is $3\times3 \times33=297$ bits, which is sufficient for hiding common personal data such as the user name, personal key and so on.

\begin{table}[t]
\label{table2}
\caption{Statistics of the number of minutiae points of each finger in FVC2002 DB1\_a}
\begin{center}
\begin{tabular}{|c|c|c|c|}
  \hline
  & minimum & maximum & average\\
  \hline
  Number of minutiae points & 11 & 92 & 49\\
  \hline
\end{tabular}
\end{center}
\end{table}

\section{Conclusions}
\label{conclusions}
In this paper, a novel scheme is proposed for data hiding in the fingerprint minutiae template. In this scheme, three strategies are proposed to maintain the accuracy of fingerprint recognition as well as the undetectability of the minutiae templates, including the bits replacement based data embedding, the optimized bits replacement base data embedding and the order preserving mechanism. In bits replacement based data embedding, the last few bits of each element in the minutiae template are replaced with the data to be hidden. This can be further improved using the optimized bits replacement based data embedding, where the replacement is performed such that its impact on fingerprint recognition is minimized. The order preserving mechanism is proposed to reduce the detectability of data hiding, which creates challenges for the attacker to identify the protected template with hidden data. We show in the experiments that the proposed scheme is able achieve sufficient capacity for hiding common personal data. Meanwhile, the performance of the fingerprint recognition is not seriously affected.
\bibliographystyle{IEEEtran}
\bibliography{Reference}
\end{document}